# Quantum measurements and high-precision control of quantum states

Yu.I. Bogdanov[*]

Valiev Institute of Physics and Technology of Russian Academy of Sciences, Moscow, Russia

**ABSTRACT**

The report presents a general approach for estimating quantum information technologies by means of fuzzy quantum measurements. The developed methods are used for precision reconstruction of quantum states under conditions of significant influence of decoherence and quantum noise. Monitoring of the amount of information about quantum states and their parameters, including the loss of information caused by the effect of quantum noise, is considered. Examples of the developed approach as applied to readout errors and phase relaxation are presented.

## 1. INTRODUCTION

It is well known that the promise of quantum computers is that some computational tasks can be performed exponentially faster on a quantum processor compared to any modern or promising classical supercomputer [1].

The great challenge is in creating a high-performance processor capable of executing quantum algorithms in an exponentially large computational space. Many different physical platforms have been proposed up to now on which quantum computing can be implemented. Some of the most promising platforms are those based on ions in traps [2, 3], systems based on atoms in traps [4,5], superconducting processors [6, 7] and photonic chips [8].

In 2021, researchers from the University of Innsbruck presented a demo for quantum computing [3]. This device fits into two 19-inch server racks and represents the world's first compact quantum computer with trapped ions that meets high quality standards [3].

Previously, in January 2019, IBM introduced the new IBM Q System One, which was the first commercial quantum computer with 20 qubits [6].

Also, in the fall of 2019, a group of scientists led by J. Martinis from Google announced the achievement of the so-called quantum supremacy, which was obtained using the Sycamore quantum processor, consisting of 53 qubits. In the article [7], published in Nature in October 2019, it was claimed that the Sycamore processor completed a task in 200 seconds, which would have taken 10 thousand years to the most powerful classical supercomputer Summit at that time. However, simultaneously with the publication of Google's results, IBM specialists attempted to refute them, arguing that in such a classical computing system as Summit, the problem under consideration could be solved in just 2.5 days or even faster [9]. The question of the degree of validity of Google's claims to achieve quantum supremacy remains a relevant topic of scientific research [10].

In December 2020, a group of Chinese scientists from the University of Science and Technology of China (USTC), in order to demonstrate quantum supremacy, implemented 76-photon bosonic sampling using the Jiuzhang photonic quantum computer [8]. The authors argued that it would take a classic modern supercomputer 600 million years of computational time to generate the number of samples their quantum processor could generate in 20 seconds.

Despite the intensive development of quantum computing for more than a quarter of a century, the creation of a full-scale quantum computer still remains an elusive dream for modern technology. At the same time, studies carried out in recent years have shown that quantum speedup is achievable in a real system and is not prohibited by any hidden physical laws. It can be stated that significant progress in the field of experimental and technological research gives rise to real hope for the creation in the medium term of quantum computing devices capable of solving practically important problems.

The achieved level of quantum information development proclaims a new era of NISQ (Noisy Intermediate Scale Quantum) technologies. Such technologies, by themselves, even before the creation of full-scale quantum computers, open up new rather broad computational possibilities, which include optimization methods, machine learning, materials science, chemistry, as well as a number of other scientific and practical fields.

It is necessary to ensure continuous improvement of NISQ - technologies so that over time they could open up opportunities to solve increasingly complex problems. To solve these problems, it is critically important to develop a system for continuous monitoring and forecasting the accuracy and efficiency characteristics of quantum information devices, depending on the degree of their integration, for computational problems of varying complexity and for various levels of decoherence and quantum noise [11, 12].

---

[*] bogdanov_yurii@inbox.ru

Along with mathematical modeling of quantum operations and algorithms, it is necessary to control them in a real experiment. The main tool for these purposes is tomography of quantum states and processes [13-16], which is designed to provide an interface between the development of the element base of quantum computers and simulators and their practical implementation. Methods of numerical analysis and statistical modeling, taking into account the effect of quantum noise, as well as the results of technological and experimental studies, allow one to give an exhaustive assessment of the quality and efficiency of the designed quantum registers, formulate requirements for experimental equipment and technology. In addition, through feedback, the developed approach allows for the best use of available resources to optimize the development of quantum information technologies.

## 2. QUANTUM TOMOGRAPHY

The quantum tomography protocol is described by the so-called instrumental matrix $X$ [15]. In the simplest case, each row $X_j$ of this matrix defines in the Hilbert space a bra-vector onto which the measured state is projected. Let the state be described by the density matrix ρ. Then the probability of registration of the corresponding event is given by the expression:

$$\lambda_j = X_j \rho X_j^+ = \text{Tr}\left(X_j^+ X_j \rho\right) = \text{Tr}\left(\Lambda_j \rho\right), \quad j = 1, ..., m, \qquad (1)$$

where $\Lambda_j = X_j^+ X_j$ is the measurement operator, m is the number of protocol rows (the total number of measured projections). In the case when the studied state is pure and described by the state vector c, expression (1) can be rewritten in another form: $\lambda_j = c^+ \Lambda_j c$.

In this paper, we restrict ourselves to protocols that reduce to the decomposition of unity. In this case, the sum of all matrices $\Lambda_j$ corresponding to the measurement operators, and being equal to $X^+ X$, is proportional to the identity matrix $I$:

$$\sum_{j=1}^{m} \Lambda_j = aI \qquad (2),$$

where $a$ is some positive constant, which can be made equal to unity by means of a simple renormalization of the measurement operators.

Under the conditions of a real experiment, it is important to achieve a high accuracy of reconstruction of a quantum state, while having at our disposal only a finite sample of size n from representatives of the quantum statistical ensemble. This goal is served by the procedures proposed in our works for the reconstruction of a quantum state using the root approach and the maximum likelihood method [15].

Methods for analyzing statistical fluctuations of reconstructed states developed in[15] make it possible to give a detailed description of the fidelity that can be achieved in various protocols of quantum measurements. Let us briefly formulate the results related to the reconstruction of pure states. Based on the complex state vector *c* and complex measurement matrices $\Lambda_j$, the corresponding real values of the doubled dimension are introduced:

$$\tilde{c} = \begin{pmatrix} \text{Re}(c) \\ \text{Im}(c) \end{pmatrix}, \quad \tilde{\Lambda}_j = \begin{pmatrix} \text{Re}(\Lambda_j) & -\text{Im}(\Lambda_j) \\ \text{Im}(\Lambda_j) & \text{Re}(\Lambda_j) \end{pmatrix}. \qquad (3)$$

The main tool for analyzing the accuracy of quantum tomography is the so-called matrix of complete information:

$$H = 2n \sum_j \frac{(\tilde{\Lambda}_j \tilde{c})(\tilde{\Lambda}_j \tilde{c})^+}{\lambda_j} \qquad (4)$$

We will assume that the sample size *n* is large enough. Then we can use the asymptotic theory of statistical estimation. In addition, the normalization condition is satisfied.

$$\langle \tilde{c} | H | \tilde{c} \rangle = 2n.$$

Let $d\tilde{c}$ be the difference between the exact and the reconstructed maximum likelihood method of the state vectors (in a real Euclidean space of doubled dimension). Then, the level of statistical fluctuations can be described by means of the chi-square distribution [15]:

$$2\langle d\tilde{c} | H | d\tilde{c} \rangle = \chi^2(\nu_H). \tag{5}$$

For the case of pure states considered here, we have $\nu_H = 2s - 1$.

Let the measurement protocol be reduced to the decomposition of the unity and be tomographically complete. Let us also specify the normalization condition in the simplest form: $\langle c | c \rangle = 1$. Then, the eigenvalues $S_{H,j}$ of the matrix *H* will make it possible to calculate the variances of fluctuations of the principal components of the reconstructed real vector of the state of the doubled dimension $\tilde{c}$:

$$d_j = \frac{1}{2S_{H,j}}, \quad j = 1, 2, \ldots, \nu, \tag{6}$$

where $\nu$ is the number of degrees of freedom of the quantum state. For a pure state in a Hilbert space of dimension *s*, the number of degrees of freedom is $\nu = 2s - 2$. Note that the matrix H is a real symmetric nonnegative definite matrix of dimension $2s \times 2s$. It has $2s$ eigenvalues, which are all non-negative. In the case of a tomographically complete protocol, one eigenvalue of the matrix H is equal to zero. It corresponds to the arbitrariness in the choice of the general phase of the state and should be discarded. In addition, it is also necessary to exclude the largest eigenvalue that meets the condition of normalization of the state. The remaining $2s - 2$ eigenvalues of the matrix *H* define the vector *d* according to (6).

The vector d is a multicomponent parameter for the universal distribution for precision, which is the generalized chi-square distribution[15]. In this case, the loss of reconstruction fidelity $1 - F$ is the following random variable:

$$1 - F = \sum_{j=1}^{\nu} d_j \xi_j^2, \tag{7}$$

where $\xi_j \sim N(0,1)$, $j = 1, \ldots, \nu$ are independent normally distributed random variables with zero mean and unit variance.

The fidelity parameter *F* in the case of pure states is $F = |\langle c^{rec} | c^{th} \rangle|^2$, where $c^{th}$ and $c^{rec}$ are the theoretical and reconstructed state vectors, respectively.

The mean and variance for the fidelity losses are determined by the following formulas:

$$\langle 1 - F \rangle = \sum_{j=1}^{\nu} d_j, \quad \sigma^2 = 2\sum_{j=1}^{\nu} d_j^2. \tag{8}$$

## 3. MODEL OF FUZZY QUANTUM MEASUREMENTS

Consider an open quantum system, the state of which at the initial moment of time is specified using the density matrix $\rho_{in}$. Being open, the system is in the process of uncontrolled exchange of information with its environment, which leads to decoherence of the state under consideration. This process can be described by an operator sum, when the state at the channel output is determined by the following formula: $\rho_{out} = \sum_k E_k \rho_{in} E_k^+$ [1, 17]. Here $E_k$ are transformation elements, or Kraus operators.

The probability of registering the event *j* during tomography of the state $\rho_{out}$, determined by formula (1), is as follows

$$\lambda_j = \mathrm{Tr}(\Lambda_j \rho_{out}) = \mathrm{Tr}\left(\Lambda_j \sum_k E_k \rho_{in} E_k^+\right) = \mathrm{Tr}(\Lambda_j^{mixed} \rho_{in}). \tag{9}$$

The operators of mixed (fuzzy) measurements are introduced here:

$$\Lambda_j^{mixed} = \sum_k E_k^+ \Lambda_j E_k. \tag{10}$$

From the point of view of the probability of event registration, tomography of the $\rho_{out}$ state by means the protocol with $\Lambda_j$ measurement operators is equivalent to tomography of the $\rho_{in}$ state by means the protocol with $\Lambda_j^{mixed}$ measurement operators.

This allows us to reconstruct the initial state of the system as it was before degradation due to interaction with environment. In other words, the decoherence channel becomes part of a generalized measurement setup that works directly with the initial state. At the same time, as will be shown below, the presence of decoherence leads to a certain degradation of the reconstruction accuracy. This degradation increases with an increase in the level of quantum noise. To compensate for the effect under consideration, it is necessary to increase the number of representatives of the statistical ensemble. Another condition for the applicability of the developed method is the need to have complete information about the characteristics of the noise in the channel (for example, you need to know the form of the operators $E_k$ or the Choi-Jamiolkowski state [1, 17-19]). To obtain this information in real experiments, one should first carry out tomography of the quantum process in the channel under consideration [16,19-21].

Let's take a closer look at decoherence caused by registration errors as an example. Kraus operators associated with registration errors are:

$$E_0 = \begin{pmatrix} \sqrt{1-p_{10}} & 0 \\ 0 & \sqrt{1-p_{01}} \end{pmatrix}, \; E_1 = \begin{pmatrix} 0 & 0 \\ \sqrt{p_{10}} & 0 \end{pmatrix}, \; E_2 = \begin{pmatrix} 0 & \sqrt{p_{01}} \\ 0 & 0 \end{pmatrix} \tag{11}$$

Here $p_{10} = P_{error}(1|0)$ is the probability of erroneous registration of state "1", when in fact there is a state "0", and $p_{01} = P_{error}(0|1)$ is the probability of erroneous registration of state "0", when in fact there is a state "1".

We assume that $|0\rangle = \begin{pmatrix} 1 \\ 0 \end{pmatrix}$ is the lower level and $|1\rangle = \begin{pmatrix} 0 \\ 1 \end{pmatrix}$ is the upper level. Then it is easy to see that: $E_1|0\rangle = \sqrt{p_{10}}|1\rangle$, $E_1|1\rangle = 0$. Thus, the operator $E_1$ specifies a "quantum leap" from the lower level $|0\rangle$ to the upper level $|1\rangle$ with probability $p_{10}$. Note that here we are not referring to a real physical "leap", but rather to the error in registering the true state of a qubit.

Similarly, the operator $E_2$ specifies a "quantum leap" from the upper level $|1\rangle$ to the lower level $|0\rangle$ with probability $p_{01}$, since $E_2|0\rangle = 0$, $E_2|1\rangle = \sqrt{p_{01}}|0\rangle$.

Now let the initial state of the qubit be a superposition of states $|0\rangle$ and $|1\rangle$ be in the form:

$$|\psi_{in}\rangle = c_0|0\rangle + c_1|1\rangle = \begin{pmatrix} c_0 \\ c_1 \end{pmatrix} \qquad (12)$$

Then, the three Kraus operators (11) define three possible alternatives for the evolution of the state:

$$|\psi_{out}^0\rangle = \begin{pmatrix} \sqrt{1-p_{10}} & 0 \\ 0 & \sqrt{1-p_{01}} \end{pmatrix} \begin{pmatrix} c_0 \\ c_1 \end{pmatrix} = \begin{pmatrix} \sqrt{1-p_{10}}\,c_0 \\ \sqrt{1-p_{01}}\,c_1 \end{pmatrix} \qquad (13)$$

$$|\psi_{out}^1\rangle = \begin{pmatrix} 0 & 0 \\ \sqrt{p_{10}} & 0 \end{pmatrix} \begin{pmatrix} c_0 \\ c_1 \end{pmatrix} = \begin{pmatrix} 0 \\ \sqrt{p_{10}}\,c_0 \end{pmatrix} \qquad (14)$$

$$|\psi_{out}^2\rangle = \begin{pmatrix} 0 & \sqrt{p_{01}} \\ 0 & 0 \end{pmatrix} \begin{pmatrix} c_0 \\ c_1 \end{pmatrix} = \begin{pmatrix} \sqrt{p_{01}}\,c_1 \\ 0 \end{pmatrix} \qquad (15)$$

From the formulas for alternatives "0" and "2" it follows that the probability of registering state $|0\rangle$ is: $P_0 = (1-p_{10})|c_0|^2 + p_{01}|c_1|^2$ (here the second term specifies the probability of erroneous registration of "zero", when in fact "one").

Similarly, from the formulas for alternatives "0" and "1" it follows that the probability of registering the state $|1\rangle$ is: $P_1 = (1-p_{01})|c_1|^2 + p_{10}|c_0|^2$ (here the second term specifies the probability of erroneous registration of "one", when in fact it is "zero").

Thus, the presented calculations show that the Kraus operators (11) set the correct values of the probabilities.

Now our task is to embed the obtained Kraus operators in the measurement operators, making them "fuzzy".

We start with the assumption that, in the ideal case, the amplitudes of the probabilities of registering states $|0\rangle$ and $|1\rangle$ are $\langle 0|\psi\rangle$ and $\langle 1|\psi\rangle$, respectively. In addition, before the measurement, the states can be subjected to some unitary transformation $U$. Then the amplitudes of the probabilities of registering states $|0\rangle$ and $|1\rangle$ will be equal to $\langle 0|U\psi\rangle$ and $\langle 1|U\psi\rangle$, respectively. Now, taking into account the registration errors, we have for the kth alternative the amplitudes $\langle 0|E_k U\psi\rangle$ and $\langle 1|E_k U\psi\rangle$, respectively. We assume that there are only $m$ alternatives, in our case $m=3$. Note that taking a set of different unitary transformations $U$, we get a set of mutually complementary amplitudes and probability distributions (in the spirit of Niels Bohr's complementarity principle).

Let us introduce the following notation:

$$\langle 0_k| = \langle 0|E_k U,\ \langle 1_k| = \langle 1|E_k U,\ k=1,...,m \qquad (16)$$

Then, the new measurement operators will be:

$$\Lambda_0 = \sum_{k=1}^{m} |0_k\rangle\langle 0_k| = \sum_{k=1}^{m} U^+ E_k^+ |0\rangle\langle 0| E_k U,\ \Lambda_1 = \sum_{k=1}^{m} |1_k\rangle\langle 1_k| = \sum_{k=1}^{m} U^+ E_k^+ |1\rangle\langle 1| E_k U \qquad (17)$$

The introduced fuzzy measurement operators $\Lambda_0$ and $\Lambda_1$ define the decomposition of unity:

Here $I$ is the identity operator

$$\Lambda_0 + \Lambda_1 = I \qquad (18)$$

The validity of (18) follows directly from the formulas $\sum_{k=1}^{m} E_k^+ E_k = I$ (preservation of the trace of the operation) and $U^+ U = I$ (unitarity of the transformation of the basis).

Let us generalize the presented formulas to the case when, along with the readout errors, there is decoherence in the channel of the quantum state transformation.

Taking into account decoherence and registration errors, we have for an individual qubit the amplitudes $\langle 0 | M_k U D_j \psi \rangle$ and $\langle 1 | M_k U D_j \psi \rangle$ for the $j$-th decoherence alternative and the $k$-th readout alternative. We assume that the considered state $|\psi\rangle$ first undergoes decoherence (Kraus operators $D_j$, $j=1,...,m_1$), then the unitary operator $U$ acts, providing a basis change, and, finally, the state undergoes reading errors during measurements (Kraus operators for measurements $M_k$, $k=1,...,m_2$). Thus, there are $m = m_1 m_2$ alternatives in total.

Let us introduce the following notation:

$$\langle 0_{jk} | = \langle 0 | M_k U D_j, \; \langle 1_{jk} | = \langle 1 | M_k U D_j, \; j=1,...,m_1, \; k=1,...,m_2$$

Then, it is obvious that a generalization of formulas (17) is:

$$\Lambda_0 = \sum_{j=1}^{m_1} \sum_{k=1}^{m_2} |0_{jk}\rangle\langle 0_{jk}| = \sum_{j=1}^{m_1} \sum_{k=1}^{m_2} D_j^+ U^+ M_k^+ |0\rangle\langle 0| M_k U D_j \qquad (19)$$

$$\Lambda_1 = \sum_{j=1}^{m_1} \sum_{k=1}^{m_2} |1_{jk}\rangle\langle 1_{jk}| = \sum_{j=1}^{m_1} \sum_{k=1}^{m_2} D_j^+ U^+ M_k^+ |1\rangle\langle 1| M_k U D_j$$

We denote by $E_{jk} = M_k U D_j$ the final Kraus operators. As a result, we get:

$$\Lambda_0 = \sum_{j=1}^{m_1} \sum_{k=1}^{m_2} E_{jk}^+ |0\rangle\langle 0| E_{jk}, \; \Lambda_1 = \sum_{j=1}^{m_1} \sum_{k=1}^{m_2} E_{jk}^+ |1\rangle\langle 1| E_{jk} \qquad (20)$$

## 4. SIMULATION RESULTS

Table 1 below shows the results of quantum tomography in the framework of the fuzzy measurement procedure for some states on the IBM quantum processor

At the first stage, the probabilities of errors $p_{01}$ and $p_{10}$ were estimated. Of the 1024 states "0", 16 were registered as "1", so the probability of an erroneous registration of 1, when it is in fact 0 is: $p_{10} = 16/1024 = 0.0156$. Similarly, out of 1024 states "1" 85 were registered as "0", so the probability of erroneous registration of 0, when it is in fact 1 is: $p_{01} = 85/1024 = 0.0830$.

At the second stage, the fuzzy measurement model was applied to the estimation of 4 states, which were the eigenvectors of the matrices $\sigma_x$ and $\sigma_y$. The sample size was $3 \cdot 1024 = 3072$ representatives (here the factor 3 corresponds to three mutually complementary probability distributions associated with the measurements of $\sigma_z$, $\sigma_x$ and $\sigma_y$, respectively).

Table 1. Comparison of the model of fuzzy quantum measurements with the standard model as applied to superconducting qubits on the IBM cloud quantum processor.

| State | Fidelity (standard model) | Fidelity (fuzzy measurement model) | Supremacy (factor showing reductions in fidelity loss) |
|---|---|---|---|
| c=[1;-i]/sqrt(2) | 0.96640 | 0.999765 | 142.7 |
| c=[1;i]/sqrt(2); | 0.99171 | 0.999275 | 11.44 |
| c=[1;1]/sqrt(2) | 0.99646 | 0.999782 | 16.26 |
| c=[1;-1]/sqrt(2); | 0.96463 | 0.998001 | 17.69 |

From the presented table, we see that the use of a fuzzy measurement model can reduce the loss of accuracy by tens, and sometimes hundreds of times.

In the case of tomographically complete and adequate quantum measurements, asymptotically for large sample sizes $n$, the average fidelity loss $\langle 1-F \rangle$ is inversely proportional to the sample size $n$, so the loss function $L = n \langle 1-F \rangle$ does not depend on the sample size.

Figure 1 shows the dependence of the fidelity loss $L$ on the dimension of the Hilbert space and the level of quantum noise caused by registration errors. Here, for each qubit, the measurements of the Pauli matrices $\sigma_z$, $\sigma_x$ and $\sigma_y$, respectively, were also considered. In the case of multi-qubit systems, measurements of the corresponding tensor products of Pauli matrices were considered. The minimum, maximum and average losses for pure states distributed according to the Haar measure are presented [22].

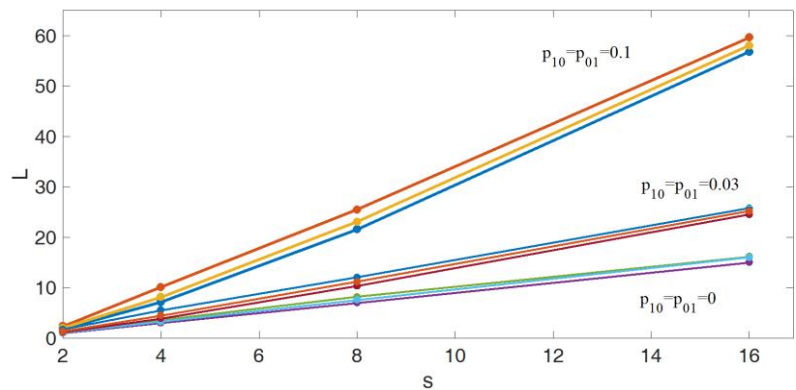

Figure 1. Dependence of the fidelity loss on the dimension of the Hilbert space and the level of quantum noise caused by registration errors.

Figure 1 shows that at the noise level $p_{01} = p_{10} = 0.03$, the losses increase by about 1.6 times compared to the ideal, and at the level $p_{01} = p_{10} = 0.1$, by 3.8 times. This means that to compensate for the loss of fidelity caused by decoherence, it is necessary to use the model of fuzzy quantum measurements and, in addition, to increase the sample sizes by 1.6 and 3.8 times, respectively.

Figure 2 shows a study of the accuracy of tomography of an 8-level qudit ($s = 8$) under conditions of phase relaxation within the framework of the fuzzy measurement model in comparison with the standard model. The tomography of pure states distributed in the Haar measure is considered. A protocol based on 9 mutually unbiased bases (MUB) was applied, the sample size in each experiment was $10^6$, the number of experiments was 200, the phase relaxation level was $g = 0.05$. When assessing the effect of phase relaxation on quantum states, we used a model that is described in detail in our work [21]. The data obtained show that the factor of reducing the fidelity loss when using the fuzzy measurement model compared to the standard was 2200, which indicates a radical advantage of the fuzzy measurement model.

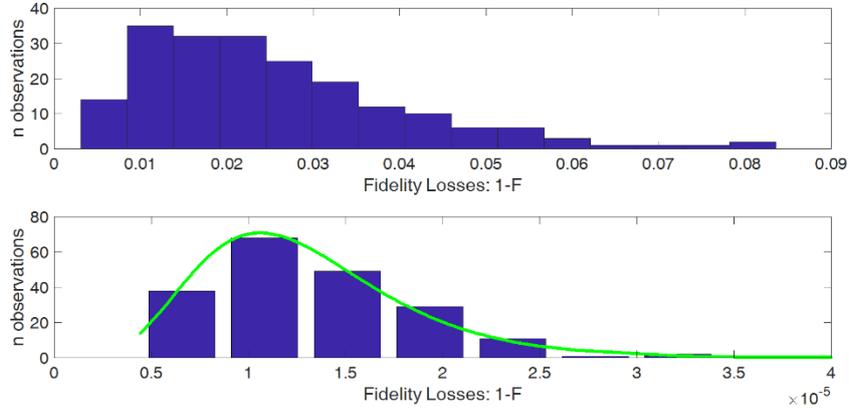

Figure 2. Investigation of the accuracy of tomography of an 8-level qudit ($s$ = 8) in conditions of phase relaxation. Comparison of standard tomography (top) with tomography within the fuzzy measurement model (bottom) is presented.

## 5. CONCLUSIONS

A general approach to estimation in quantum information technologies using fuzzy quantum measurements has been developed. Using examples with superconducting qubits on an IBM cloud quantum processor, it is shown that the use of a fuzzy measurement model can reduce the loss of fidelity in reconstructing quantum states by tens, and sometimes hundreds of times. Numerical examples of studies of decoherence processes caused by registration errors and phase relaxation, carried out for systems with Hilbert space dimensions from 2 to 16 and for various levels of quantum noise, have shown the radical superiority of the model of fuzzy quantum measurements in comparison with the standard model. The research results are of practical importance for the problems of ensuring the quality and efficiency of quantum information technologies.

## ACKNOWLEDGMENTS

This work was supported by the Ministry of Science and Higher Education of the Russian Federation (program no. FFNN-2022-0016 for the Valiev Institute of Physics and Technology, Russian Academy of Sciences), and by the Foundation for the Advancement of Theoretical Physics and Mathematics BASIS (project no. 20-1-1-34-1)